# III-V Tri-Gate Quantum Well MOSFET: Quantum Ballistic Simulation Study for 10nm Technology and Beyond

*Kanak Datta and Quazi D. M. Khosru
Department of Electrical and Electronic Engineering
Bangladesh University of Engineering and Technology
Dhaka-1000, Bangladesh
*Phone No.: +8801733653674; Email: kanakeee08@gmail.com

**Abstract:**

*In this work, quantum ballistic simulation study of a III-V tri-gate MOSFET has been presented. At the same time, effects of device parameter variation on ballistic, subthrshold and short channel performance is observed and presented. The ballistic simulation result has also been used to observe the electrostatic performance and Capacitance-Voltage characteristics of the device. With constant urge to keep in pace with Moore's law as well as aggressive scaling and device operation reaching near ballistic limit, a full quantum transport study at 10nm gate length is necessary. Our simulation reveals an increase in device drain current with increasing channel cross-section. However short channel performance and subthreshold performance get degraded with channel cross-section increment. Increasing device cross-section lowers threshold voltage of the device. The effect of gate oxide thickness on ballistic device performance is also observed. Increase in top gate oxide thickness affects device performance only upto a certain value. The thickness of the top gate oxide however shows no apparent effect on device threshold voltage. The ballistic simulation study has been further used to extract ballistic injection velocity of the carrier and ballistic carrier mobility in the channel. The effect of device dimension and gate oxide thickness on ballistic velocity and effective carrier mobility is also presented.*

## 1. Introduction:

With Si CMOS technology gradually reaching its ultimate limit, integration of III-V materials in multigate device structures has become a topic of intensive research and study. With excellent carrier transport properties and lower effective mass, III-V semiconductors outperform Si in many aspects of nanoscale device applications [1]. With sub-10nm node technology approaching fast, counteracting short channel effects (SCEs), sub-threshold conduction, improving electrostatic behavior, reduced source/drain tunneling and gate leakage current have become aspects of research and investigation. Multigate device architectures like FinFETs, GAA FETs provide better control over channel carrier accumulation and improved short channel performance over planar structures. Integration of III-V semiconductors with multigate device architectures can certainly meet the demands of future technology nodes [2]. First experimental demonstration of III-V semiconductors in non-planar tri-gate architecture took place in 2009[3]. Since then, various III-V semiconductors and their alloys have been employed successfully in multigate architectures like FinFET, Tri-Gate, GAAFETs with increased scaling and improved short channel performance[3–10]. A sub-100nm III-V tri-gate device that incorporates bi-layer high-k dielectric material ($Al_2O_3/HfO_2$ with EOT<1 nm) with fin width scaled down to *30nm* has been demonstrated and reported in recent literature[10]. With the implementation of *14nm* FDSOI (Fully Depleted Silicon on Insulator) technology in recent times, III-V multigate devices could be the ultimate successors of low power CMOS applications in future technology nodes. Therefore, a detailed simulation study of transport characteristics and the effect of various parameters on carrier transport phenomenon in sub-*10nm* region operations are required to fully understand the potential of these devices. In this work, quantum ballistic simulation study of a III-V tri-Gate MOSFET with similar architecture as in [10] with *10nm* gate length, using self-consistent analysis and NEGF formalism is presented. Carrier density obtained from ballistic simulation in the channel has been used to extract the Capacitance-Voltage (C-V) characteristics of the device. At the same time, effects of various parameters like channel material, channel dimension, gate oxide thickness on the ballistic device performance have been observed. The effect of top gate oxide thickness on carrier transport is observed and the critical top gate oxide thickness for carrier transport in *10nm* regime operation is reported. At the same time, effects of device dimension on ballistic carrier velocity and ballistic mobility is observed and presented. Effects of scattering due to impurity, electron-phonon interaction and interface roughness have not been considered in this study. In III-V devices, surface roughness scattering affects device performance significantly by lowering carrier velocity in the channel and consequently degrading device 'on' current. Presence of trapped charges at the high-k oxide and semiconductor interface could introduce Fermi level pinning and degrade device performance significantly. Recent studies have shown, the effect of surface roughness scattering becomes lower at wider channel devices [11]. However, as device dimensions are scaled, surface roughness becomes an important issue and degrades device mobility in InGaAs material system. Apart from surface roughness scattering, alloy disorder scattering, impurity scattering, phonon scattering also degrades device performance in III-V materials[12]. In NEGF formalism, the effects of both elastic and inelastic scattering events can be taken into account in ballistic simulation using Büttiker probe method mentioned in [13,14]. Inclusion of non-ideal scattering would decrease effective mobility in the channel, lower carrier velocity, increase channel resistance and degrade device performance.

**2. Device Structure:**
Fig. 1 shows a simplified structure of the device used in this study. Here, $W_{Fin}$ refers to the width of the channel and $H_{Fin}$ refers to the height of the channel. $L_G$ refers to the gate length of the device. The simulated device structure incorporates *2nm* high-k $HfO_2$ as gate dielectric, high mobility $In_{0.7}Ga_{0.3}As$ channel material, $In_{0.52}Al_{0.48}As$ as back barrier layer. Gate length is kept fixed at *10nm*. In our simulation, we have considered source/drain extension region to be 10nm each. Here, in the figure, $L_{SE}$ and $L_{DE}$ refers to the source and drain extension regions respectively. $W_{Fin}$ and $H_{Fin}$ are considered as the dimensional parameters of the channel. Metal workfunction has been considered to be *4.8 eV*.

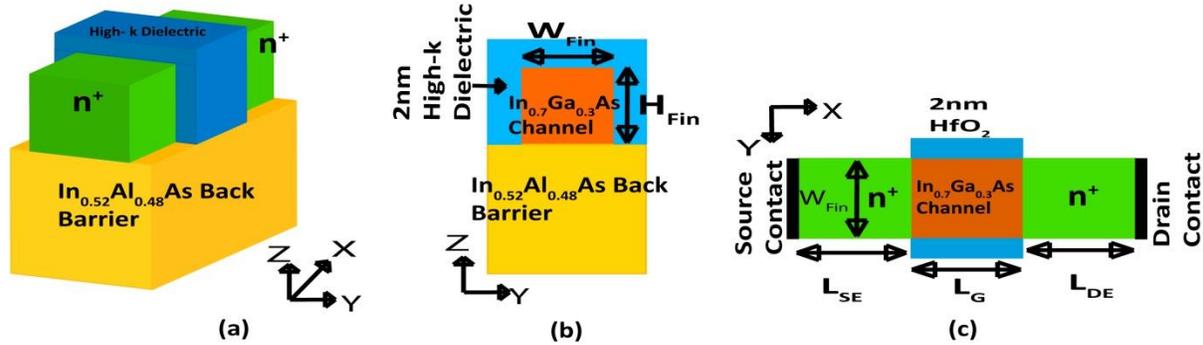

Fig. 1. Device structure used in this study: (a) Simplified 3D view; (b) A 2D cross-section of the device; (c) Cross-section showing the channel and source/drain extension regions.

**3. Device Simulator Design and Implementation:**

Simulation of transport characteristic based on ab-inito approach using Density Functional Theory (DFT) can be found in literature [15]. Although several computer aided software packages are available, these approaches are computationally extensive and have some inherent fundamental limitations such as bandgap underestimation. Huge computational burden prohibits the ab-initio approach to be used as useful tool for simulating large scale systems. Empirical tight binding approach is an way around this computational burden however it has its limitations as well [16]. This method is based on tight binding parameterization of the structure and these parameters vary depending on atomic configurations and structures. However, this model allows simulation of large device structures and systems.
In this study, we have used effective mass Hamiltonian along with fast uncoupled mode space (FUMS) approach for transport analysis of the device. The process is a bit more simplified and computationally efficient. With proper choice of device parameters, this approach can match experimental results with reasonable accuracy[17], [18]. Therefore, we have opted for a simplified fast uncoupled mode space approach with effective mass Hamiltonian. In recent years, there has been extensive research work on the transport analysis of the III-V multigate device structures [19–21]. In this work, we have implemented the simplified quantum transport simulator using self-consistent analysis incorporating NEGF formalism. The self-consistent simulation approach is implemented using COMSOL Multiphysics [22] and MATLAB. The simulation approach has already been implemented for analysis of multigate planar [23,24] and non-planar device architectures [13]. The developed simulator has also been validated with reported results available in [13].
Liu et. al explored the validity of simplistic parabolic conduction band model with effective mass approximation and concluded that, for UTB Si device structures, simple parabolic conduction band with bulk effective mass approximation gives reasonably good results for device channel thickness up to *3nm* [17,25]. However, for deeply scaled UTB III-V devices, due to strong quantum confinement, conduction band non-parabolicity can affect confinement effective masses significantly and change contribution of different valleys. A simplified parabolic conduction band model with bulk effective mass approximation fails to take into account these effects and therefore, could lead to erroneous results in III-V UTB devices. Liu et. al [26] reported that in UTB III-V devices, $\Gamma$ valley shows lowest energy and significant carrier population according to tight-binding (TB) model. In our simulation, we have considered only $\Gamma$ valley for carrier transport and used $\Gamma$ valley effective masses for UTB InAs and GaAs films reported in [26]. For basic transport analysis, we have considered width ($W_{Fin}$) and height of the channel ($H_{Fin}$) to be *6 nm*. High-k $HfO_2$ has been used as the gate dielectric material. The dielectric thickness is kept at *2nm* (*EOT<1nm*). The thickness of the back barrier $In_{0.52}Al_{0.48}As$ layer has been kept fixed at *20nm*. The source and drain extension doping is kept fixed at $0.8 \times 10^{26}$ m$^{-3}$. Bottom $In_{0.52}Al_{0.48}As$ buffer layer is lattice matched with InGaAs material system at 0.53 In mole fraction. In our simulation, we have varied the mole fraction of In in the InGaAs quantum well. Variation of In mole fraction results in a lattice constant different from bottom buffer layer and strain is developed. Strain modulates and changes the conduction and valance band edges at heterostructure interfaces. In this work, model solid theory has been used to calculate the conduction band offsets at heterostructure interfaces taking into account the effect of strain [27,28]. The effect of strain on carrier effective mass has been considered negligible. In order to simulate the device in quantum ballistic regime, we have used the self-consistent simulation procedure depicted in [13]. Assuming an initial charge density profile, 3D Poisson equation is solved in Finite Element Method (FEM) using COMSOL Multiphysics [22] to extract the potential profile $E_C(x,y,z)$ inside the device. To extract the subband profile and carrier density profile inside the device structure, solution of 3D Schrodinger equation is required. Application of mode space approach reduces this 3D problem into two separate lower dimensional problems. In Uncoupled Mode Space (UMS) approach, Schrodinger equation must be solved at every slice of the device structure along gate confinement direction to obtain subband profile, which can be used to obtain the overall carrier density using NEGF formalism. However, a more efficient subband calculation is possible through the application of Fast Uncoupled Mode Space (FUMS) approach. In this approach, the wavefunction for a particular eigen energy is considered to be invariant in the channel along the transport direction (X direction in Fig. 1) of the device. Which means, carrier wavefunctions and subband energies can be calculated by solving Schrodinger equation for a single 2D slice in the channel.

From the conduction band profile extracted using Poisson equation, we calculate an average conduction band profile:

$$\bar{E}_C(y,z) = \frac{1}{L_x} \int_0^{L_x} E_C(x,y,z)dx \quad (3.1)$$

Using this average conduction band profile, 2D Schrodinger equation is solved to extract the subband profile in the channel and carrier wave functions [13]:

$$[-\frac{\hbar^2}{2}\frac{\partial}{\partial y}(\frac{1}{m_y^*(y,z)}\frac{\partial}{\partial y}) - \frac{\hbar^2}{2}\frac{\partial}{\partial z}(\frac{1}{m_z^*(y,z)}\frac{\partial}{\partial z}) + \bar{E}_C(y,z)]\bar{\psi}^m(y,z) = \bar{E}_{sub}^m \cdot \bar{\psi}^m(y,z) \quad (3.2)$$

Here, $\bar{E}_{sub}^m$ refers to the eigen energy of the $m^{th}$ subband in the channel. After calculating average subband energy, we calculate the overall subband profile in the channel using first order perturbation theory using the following equation [13]:

$$E_{sub}^m(x) = \bar{E}_{sub}^m + \int_{y,z} E_C(x,y,z)|\bar{\psi}^m(y,z)|^2 \, dydz - \int_{y,z} \bar{E}_C(y,z)|\bar{\psi}^m(y,z)|^2 \, dydz \quad (3.3)$$

Using the subband energies in the channel, the electron density is found using NEGF approach [23]. The retarded Green's function can be written as [29]:

$$G(E) = [EI - H - \Sigma_S(E) - \Sigma_D(E)]^{-1} \quad (3.4)$$

Where, $H$ refers to the device Hamiltonian formulated using the subband energy profiles in the channel. $\Sigma_S(E)$ ($\Sigma_D(E)$) refers to self-energy matrices corresponding to the coupling between channel and source (drain) reservoirs [30]. Self-energy matrices allow us to work within the simulation region presented in fig. 1 without including infinite reservoirs i.e. source and drain. The effect of infinite source and drain reservoirs are taken into account by self-energy matrices. From the self-energy matrices and retarded Green's function, the spectral density functions for source ($A_S$) and drain ($A_D$) contacts are calculated:

$$A_S = G\Gamma_S G^\dagger \quad \text{and} \quad A_D = G\Gamma_D G^\dagger \tag{3.5}$$

Here, $\Gamma_S$ and $\Gamma_D$ refers to the broadening matrices of source and drain contacts respectively [30], which are formulated as:

$$\Gamma_S = i(\Sigma_S - \Sigma_S^\dagger) \text{ and } \Gamma_D = i(\Sigma_D - \Sigma_D^\dagger) \tag{3.6}$$

From spectral density functions, 1D carrier density for $m^{th}$ subband in the channel can be formulated as [13]:

$$n_{1D}^m = \frac{1}{\pi a} \int_{-\infty}^{\infty} (A_S^m f(E,\mu_S) + A_D^m f(E,\mu_D)) dE \tag{3.7}$$

Where, $a$ refers to the grid spacing in the channel direction (X direction). Afterwards, 3D carrier density for the $m^{th}$ subband is calculated from 1D carrier density and average carrier wavefunction using the following equation:

$$n_{3D}^m(x,y,z) = n_{1D}^m(x) |\bar{\psi}^m(y,z)|^2 \tag{3.8}$$

Total 3D carrier density is calculated taking into account the overall contribution of all the subbands. After the calculation of 3D carrier density, 3D Poisson equation is solved to extract the new potential profile. Then using an update co-efficient, potential profile is updated and the updated potential profile is used for further calculation in self-consistent procedure. Once self-consistency is achieved, current for a subband is calculated using NEGF formalism using the following equations [30]:

$$T^m(E) = Trace(\Gamma_S(E) G^m(E) \Gamma_D(E) G^{m\dagger}(E)) \tag{3.9}$$

$$I_{SD}^m = \frac{q}{\pi \hbar} \int_{-\infty}^{\infty} T^m(E)[f_S(E,\mu_S) - f_D(E,\mu_D)] dE \tag{3.10}$$

Here, $T(E)$ refers to the transmission co-efficient. $I_{SD}$ refers to the drain current. $f_S$ and $f_D$ refer to the Fermi function for source and drain. $\mu_S$ and $\mu_D$ refer to source and drain fermi levels. To calculate the total device current, we need to add up the contributions from all the subbands. In this work, source fermi level is kept fixed at *0.0V* for all cases.

### 4. Results and Discussion:

### 4.1 Simulator Validation:

In this study, simulation process depicted in [13] has been used. Fig. 2a shows the Id-Vg characteristics of a rectangular Si NWFET obtained from the developed simulator along with the results reported in [13]. For this analysis, the width and height of the Si fin has been considered to be *4nm*. The channel is considered to be <101> oriented. The effective masses used in the Si NWFET analysis have been calculated using the formulation used in [31]. In this case *1nm* $SiO_2$ has been used as gate dielectric material. Fig. 2b shows the first subband energy in the channel along with 1D carrier density obtained from UMS and FUMS approaches for InGaAs tri-gate device. The figure illustrates that, efficient FUMS approach captures the subband energies and carrier densities in the channel without any significant loss of accuracy. For other higher subbands (not shown in Fig. 2b) FUMS and UMS approaches show negligible difference in subband energy and carrier density.

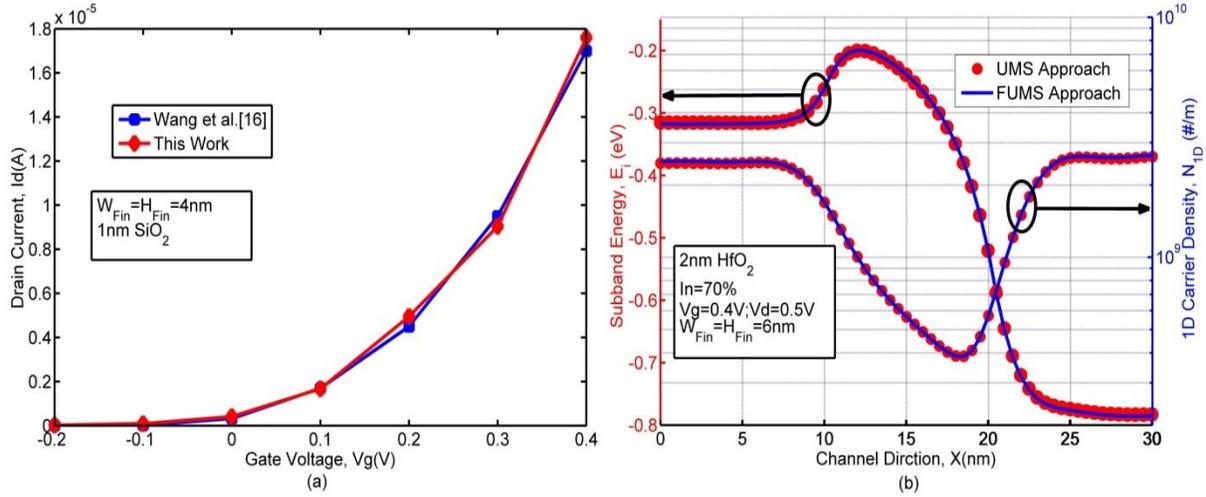

Fig. 2. (a) Id-Vg characteristics for a Si rectangular NW transistor using the developed simulator along with the results reported in [13]. (b) Validation of FUMS approach. 1D carrier density for the first subband along with subband energy along the channel at $Vg=0.4V$ and $Vd=0.5V$ for the III-V tri-gate MOSFET with $W_{Fin}=H_{Fin}=6nm$ and 0.7 In mole fraction in the channel. In all these cases, source fermi level has been kept fixed at $0.0V$.

**4.2 Transport Analysis of Tri-gate InGAaAs QW MOSFET at 0.7 In Mole Fraction:**

From transmission point of view, nanoscale transistors can be visualized by a potential barrier between source and drain. The height of the barrier is ideally modulated by the gate voltage. Carriers are injected into the channel from thermal equilibrium reservoirs (source and drain). The carrier density in the channel is governed by electrostatic effects exerted by gate and drain contacts when channel lengths are small. This simplified model of ballistic nanotransistor also allows us to get an idea of average carrier velocity in the channel. According to Natori et al. [32], the maximum average carrier velocity at the beginning of the channel (top of the barrier region) is the equilibrium thermal velocity. However, above threshold voltage, the average carrier velocity near the source end can go above thermal velocity [33]. The average carrier velocity is also affected by drain voltage. At low drain bias, the average carrier velocity remains close to zero. This can be explained by the injection of negative velocity carriers from the drain contact. However, as the drain bias increases, the potential barrier for the drain injected carriers increases and this leads to suppression of drain injected carriers. Therefore average carrier velocity at the source end (top of the barrier) increases with increased drain bias and saturates gradually. However, maximum average carrier velocity along the channel increases and the position in the channel where the carriers achieve maximum velocity gradually shifts towards the drain end [33]. In nanowire transistors, total drain current at 'on' condition can be written as [34]:

$$I_D = qn_{ToB}v_{ToB} \qquad [4.2.1]$$

Here, $q$ refers to electron charge, $n_{ToB}$ refers to the top of the barrier carrier density and $v_{ToB}$ refers to the average carrier velocity at the top of the barrier. Using this equation, we can calculate the average carrier velocity from our simulation results.

Fig. 3a shows the average carrier velocity at different positions in the channel at 0.7 In mole fraction at different drain bias conditions. Average carrier velocity increases as we gradually move toward the drain along the channel due to increased effect of drain electric field. The inset figure shows $1^{st}$ subbband energy along with 1D carrier density in the channel at Vg=0.4V for three drain bias voltage conditions. As seen from the inset figure, the energy barrier height near the source end gets modulated by drain electric field which allows carrier density at the top of the barrier to increase at high drain voltages. The 1D carrier density profile shows a gradual lowering from the source end to drain end of the channel. This can be explained by the decrease in drain injected carriers at high drain bias voltage condition. Carrier density in the channel includes contribution from both drain and source. With increased drain voltage, drain injected carriers see a higher potential barrier and therefore carrier injection from drain is suppressed. This leads to a gradual lowering in 1D carrier profile from source to drain. Fig. 3b shows the ballistic injection velocity (average carrier velocity at the top of the barrier) and 1D carrier density at the top of the barrier at different drain bias conditions. We see a gradual increase in the injection velocity and 1D carrier density with increased drain bias voltage. The increase in ballistic injection velocity can be attributed to the suppression of opposite velocity drain injected carriers as drain bias voltage is increased and the top of the barrier carrier density being populated by the carriers injected from source [33]. The increment in 1D carrier density in fig. 3b at the top of the barrier can be attributed to the gradual lowering of the top of the barrier energy with increased drain voltage due to 2D electrostatics.

Fig. 4a shows the 1D carrier density in the channel at the top of the barrier as a function of gate voltage at 2 different drain voltages for two different In compositions. From this figure, we can observe that, 1D carrier density at $Vd=0.0V$ is higher for higher In composition. The figure also reveals a lower threshold voltage for higher In composition channel which can be explained by the lowering of bandgap with increasing In composition in InGaAs material system [35]. The figure also shows top of the barrier carrier density at $Vd=0.5V$. Due to gradual lowering of top of the barrier energy with increased drain bias and short channel effects we see an increase in top of the barrier carrier density compared to the condition at $Vd=0.0V$. These additional carriers are injected from source and are not balanced by the gate electrostatics. The inset figure shows C-V characteristics of the device at two different In mole fractions. For the extraction of ballistic current, only first two subbands have been considered in the simulation. Fig. 4b shows the overall transmission co-efficient calculated at $Vg=0.4V$ and $Vd=0.5V$. Transmission co-efficient has been calculated using Eq. (3.9) for each subband individually. For ballistic transport, the transmission co-efficient for the carriers in a subband can reach a maximum value of 1. For two subbands we get an overall maximum value of 2 and the step in the transmission co-efficient profile occurs at about the top of the barrier energy of the subbands.

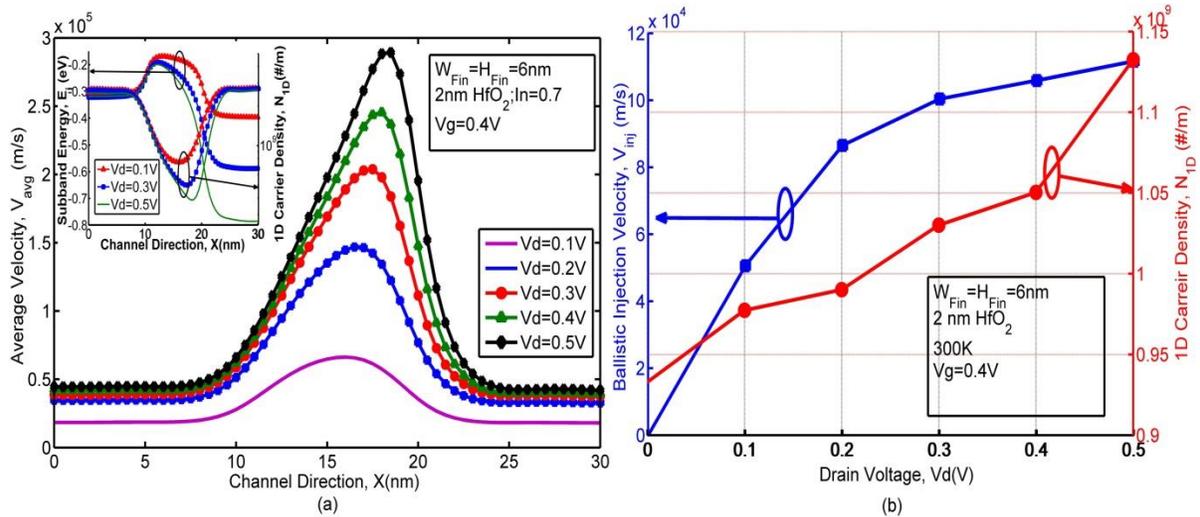

Fig. 3. (a) Average carrier velocity in the channel at different drain bias voltage at $Vg=0.4V$. Average carrier velocity increases as we move towards the drain which is due to strong velocity overshoot at higher drain electric field. The inset figure shows the subband energy as a function of drain bias voltage at $Vg=0.4V$. Slight energy barrier lowering can be observed with increased drain bias voltage. (b) Ballistic injection velocity i.e. average carrier velocity at the top of the barrier and 1D carrier density at the top of the barrier at different drain bias voltage conditions at $Vg=0.4V$.

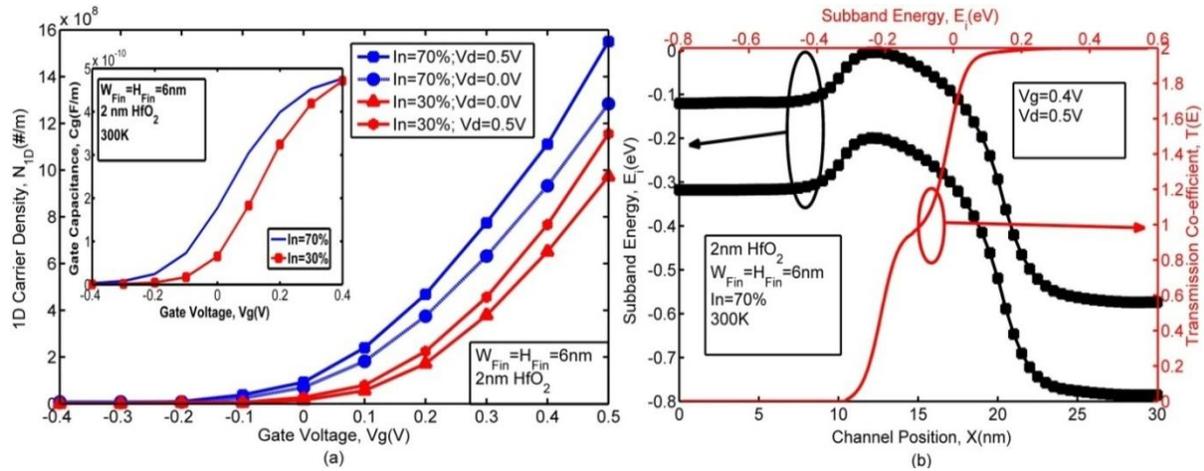

Fig. 4. (a) 1D carrier density in the channel at the top of the barrier at different gate voltages for two different drain bias conditions. Carrier density at the top of the barrier increases with increased drain voltage. Channel with lower In mole fraction shows higher threshold voltage due to higher bandgap. The figure also reveals a lowering of threshold voltage with drain voltage increment which is due to drain induced barrier lowering (DIBL) effect. The inset figure shows the C-V characteristics of the device. (b) Transmission co-efficient along with subband energies for two subbands in the channel. As ballistic transport is assumed, the overall transmission co-efficient reaches a value of 2 for two subbands shown in the figure.

Fig. 5a shows the energy resolved current density for the first subband. It basically shows the current contribution of different energy levels in the energy grid used in NEGF formulation. The figure reveals that although most of the carriers with energy above the top of the barrier energy contribute to overall drain current, some under the barrier transport is also present. This may be attributed to under the barrier quantum tunneling that becomes significant at very low channel length. From energy resolved current density plot, we can get a quantitative idea of source-to-drain tunneling contribution to total device current. From our analysis, the total contribution of source-to-drain tunneling to total drain current at 'on' condition was found to be approximately 8%. Fig. 5b shows the Id-Vg characteristics of the device in both linear and log scale at two different drain bias voltages, $V_d=0.1V$ and $V_d=0.5V$. The logId-Vg characteristics can be used to extract important device parameters like Subthreshold Swing (SS) and Drain Induced Barrier Lowering (DIBL). From our simulation, we have obtained the SS to be *93 mV/dec* and DIBL value of *100.5 mV/V* at *Vd=0.5V* for 0.7 In mole fraction in the channel.

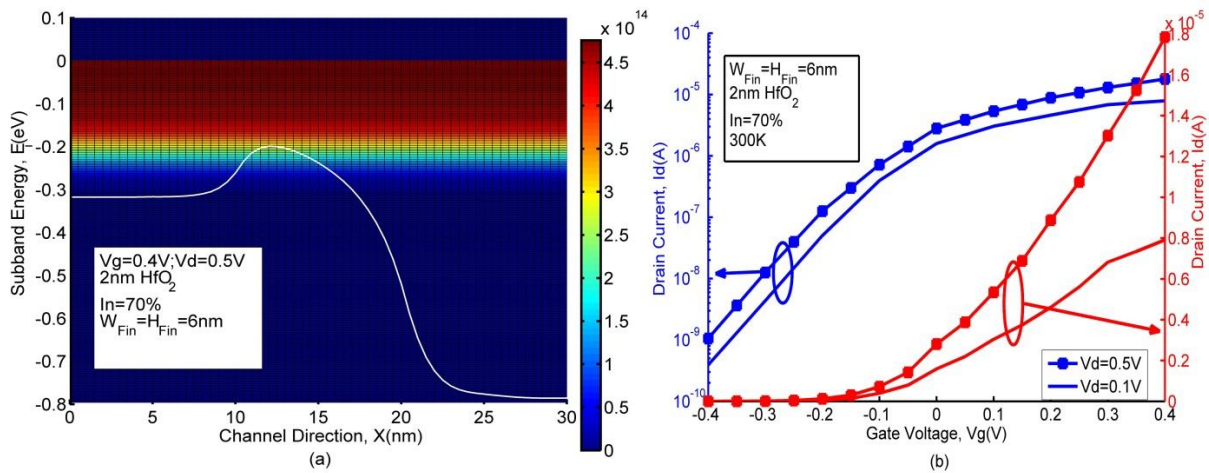

Fig. 5. (a) Energy resolved current density (A/eV) for the first subband for the 10nm device. Although significant carrier transport occurs at energies above the top of the barrier, states having energies lower than the top of the barrier marginally contribute to the overall carrier transport as a result of quantum tunneling. (b) Id-Vg

characteristics in both linear and log scale at two different drain bias voltage conditions. The logId-Vg curve at $Vd=0.5V$ has been used for the calculation of SS.

### 4.3 Effect of Variable In mole Fraction in the Channel:

In our study, we have changed the composition of In in the channel and observed the effect of variable In composition on the transport characteristics. The In composition in the InAlAs bottom barrier layer has been kept fixed at 52%.

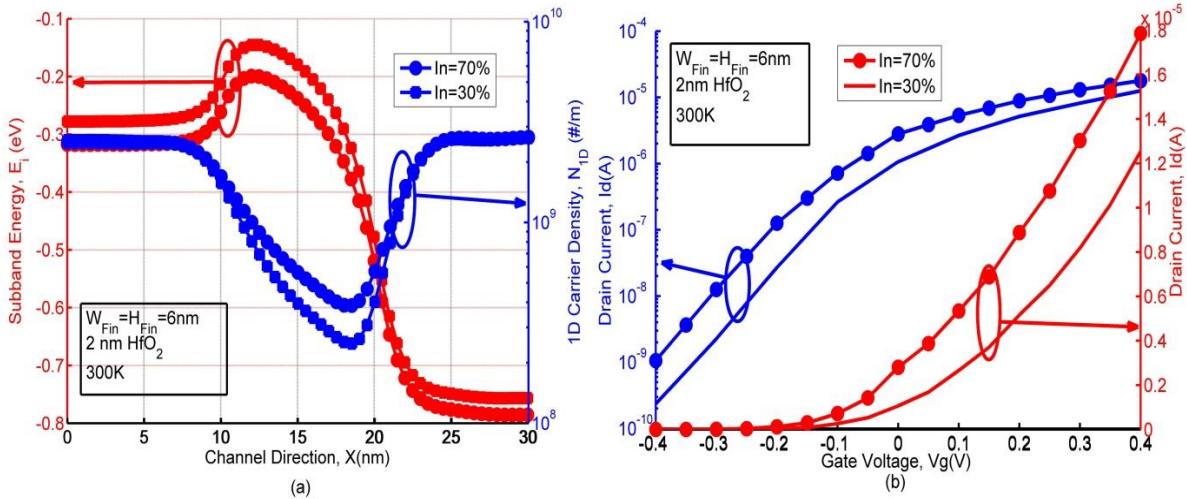

Fig. 6. (a) 1st subband energy along with 1D carrier density for two different In mole fractions in the channel at $Vg=0.4V$ and $Vd=0.5V$. Higher In composition results in lower subband energies and a slightly lower energy barrier height between source and channel. (b) Id-Vg characteristics at two different In mole fractions in the channel at $Vd=0.5V$. Higher drain current at higher In composition can be attributed to lower effective mass and higher carrier velocities. Lowering of threshold voltage is also observed at higher In mole fraction.

Fig. 6a shows the 1st subband energy and 1D carrier density in the channel for two In compositions at $Vg=0.4V$ and $Vd=0.5V$. The figure illustrates lowering of subband energies with increasing In composition in the channel. At the same time a slight lowering of the energy barrier height can also be observed from the figure at high In composition. This can be attributed to the lowering of energy bandgap with increasing In composition in the channel. The top of the barrier energy appears to be occurring at the same position in the channel. However, due to lower subband energies, source injected carriers appears to be higher for higher In composition and carrier density will be higher. Fig. 6b shows the Id-Vg characteristics for the simulated device at $Vd=0.5V$ in both linear and log scale. Device with higher In composition leads to higher drain current. Variation in In composition results in a change in threshold voltage as well. The variation of threshold voltage with varying In composition and channel dimension is shown in a later section.

### 4.4 Effect of Gate Oxide Thickness:
We have explored the effect of gate oxide thickness on the transport characteristics of the tri-gate device. The effects on carrier density, subband energy, device on current have same observed and reported. We have also reported the critical top gate oxide thickness in the ballistic regime.
Fig. 7a shows the effect of gate oxide thickness on subband energies and carrier density. We have shown only first subband in this figure. The figure shows subband energies for 2 gate oxide thickness conditions: *2nm* and *4nm*. With increasing oxide thickness, we see a lowering of subband energies and decrement in 1D carrier density which underlines lowering gate electrostatic effects with increased oxide thickness. Fig. 7b shows the effect of gate oxide thickness on 1D carrier density as a function of gate bias voltage at $Vd=0.0V$. The inset figure reveals the C-V characteristics at different gate oxide thickness conditions. With increased oxide thickness, the gate electrostatic effect gets lowered and therefore we see a decrement in the gate capacitance.

Fig. 8a shows the effect of top gate oxide thickness variation on 1$^{st}$ subband energy. The thickness of the sidewall oxide has been kept fixed at *2nm*. The figure also shows 1D carrier density in the channel. From the figure we can see that top of the barrier energy gets slightly lowered as we decrease top gate oxide thickness. However, the energy barrier lowering with top gate oxide thickness reduction seems to be very low with compared to fig. 7a which underlines strong electrostatic control of sidewall gates on device operation in tri-gate architecture. 1D carrier density in the channel is also slightly higher at lower oxide thickness condition. Fig. 8b shows the effect of top gate oxide thickness on the 1D carrier density at the top of the barrier at *Vd=0.0V* at different gate bias voltages. The inset figure shows the C-V characteristics at *Vd=0.0V*. Device with lower top gate oxide thickness shows slightly better control over channel electrostatics and therefore a slightly higher gate capacitance. In all these cases, we have considered a rectangular fin of $W_{Fin}=H_{Fin}=6nm$. The variation of gate capacitance with only top gate oxide thickness increment appears to be small compared to the case of gate oxide thickness variation on all sides which reveals strong electrostatic effect of the side-wall gates in a tri-gate architecture.

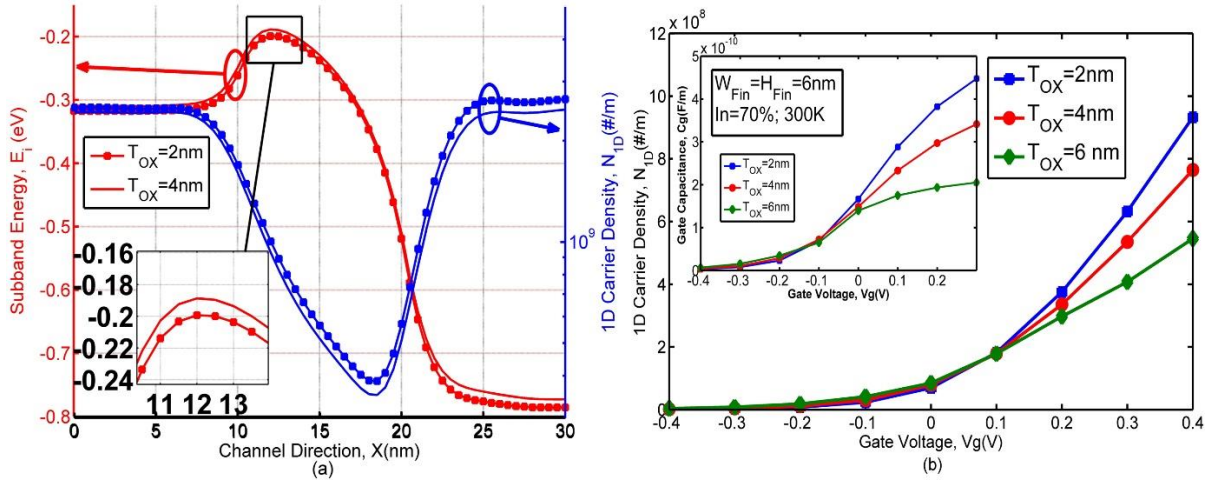

Fig. 7.(a) 1D carrier density along with first subband energy in the channel for the simulated device at *Vg=0.4V* and *Vd=0.5V* at two different gate oxide thickness conditions. The zoomed portion shows slight difference in subband energies. (b) Carrier density at the top of the barrier at *Vd=0.0V* at different gate bias voltages. Increment is oxide thickness lowers carrier density at the top of the barrier position and results in a lower gate capacitance at shown in the inset figure.

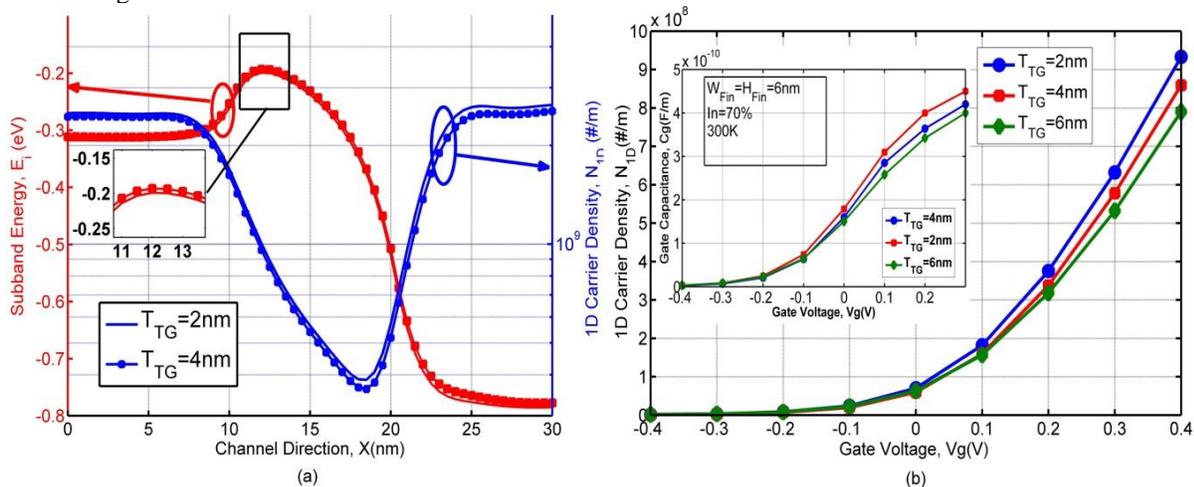

Fig. 8.(a) 1D carrier density along with first subband energy in the channel for the simulated device at *Vg=0.4V* and *Vd=0.5V* at two different top gate oxide thickness conditions. Difference in subband energies reveal lowered electrostatic effect of the top gate. (b) Carrier density at the top of the barrier at *Vd=0.0V* at different top gate oxide thickness conditions. The inset figure reveals the C-V characteristics at different top gate oxide thickness conditions.

Raisul et. al presented a distinction between double gate (DG) and tri-gate (TG) FinFETs on the basis of electrostatic and transport analysis in terms of critical oxide thickness [36]. For SOI FinFETs, after a certain thickness of the top gate oxide the device 'on' current appears to be independent of top gate oxide thickness. We have observed similar effects in our study. Fig. 9a shows the variation of device 'on' current and 1D carrier density at device 'on' condition with top gate oxide thickness. As mentioned before, we define $Vg=0.4V$ and $Vd=0.5V$ as device 'on' condition. With increased top gate oxide thickness we see a decrease in device 'on' current. However, the variation of 'on' current appears to be gradually decreasing with increased oxide thickness. This phenomenon indicates the effect of top gate gradually becoming trivial and indicates strong effect of the sidewall gates on carrier transport and current drive of the device. The lowering of 1D carrier density with top gate oxide thickness also reveals lowered electrostatic control of the top gate. The simulation reveals negligible effect on the threshold voltage as we increase top gate oxide thickness which underlines the strong effect of the sidewall gates on device threshold. From these results, the critical oxide thickness for the tri-gate device with square channel cross-section can be reported in the region of *7nm-8nm*. Fig. 9b shows the 'on' current and carrier density variation when gate oxide thickness is varied on all sides. Unlike the case of top gate oxide thickness variation, the simulation reveals gradually lowering electrostatic effect of both top and side-wall gates. As a result, 1D carrier density and 'on' current both decease as we increase gate oxide thickness on all three sides.

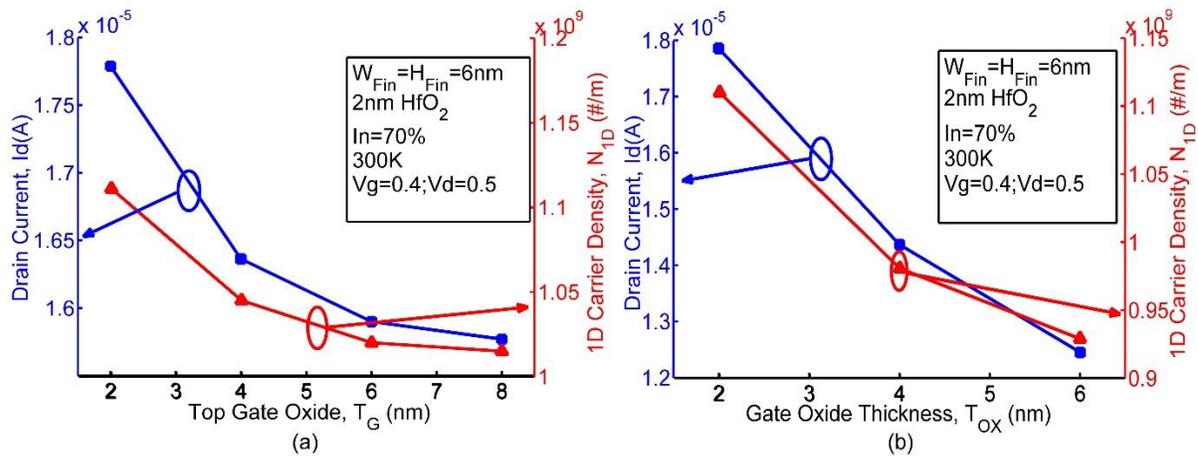

Fig.9. (a) Variation of on current and carrier density with top gate oxide thickness conditions at $Vd=0.5V$ and $Vg=0.4V$. In composition in the channel is kept at 0.7. The variation of 'on' current gradually decreases with increased top oxide thickness as the device approaches critical top gate oxide thickness limit. Carrier density at the top of the energy barrier gradually decreases with increased top gate oxide thickness at 'on' condition of the device. (b) The variation of device 'on' current and carrier density when gate oxide thickness is varied on all sides. The simulation reveals lowering electrostatic effect of all three gates with increased oxide thickness.

**4.5 Effect of Channel Dimension:**

In our basic simulation, we have considered square cross-section fin for transport analysis. However, we have also explored the effect of rectangular fin dimension on transport characteristics. We have kept the fin width ($W_{Fin}$) fixed at *6nm* and varied the height of the fin from *6nm* to *14nm* in 4 nm steps and observed the effects. We have also observed the effect of square channel ($W_{Fin}=H_{Fin}$) cross-section on transport characteristics of the device. In these cases, all other device parameters have been kept unchanged.
Fig. 10a shows first subband energy and carrier density at two different fin heights at $Vg=0.4V$ and $Vd=0.5V$. As we increase the fin height, the overall cross-section of the device increases and therefore we see a lowering in the subband energies. At the same time, as carrier concentration in the channel is also controlled by source and drain fermi level, due to lowering of the subband energies, we observe an increase in 1D carrier concentration. Only first subband carrier concentration is being shown here. Fig. 10b shows the top of the barrier carrier density at $Vd=0.0V$ for different gate voltage conditions. As mentioned, increasing fin height increases overall cross-section of the device, lowers eigen energies in the quantum well and therefore more carriers get to accumulate in the channel. The inset figure also shows the C-V characteristics of the device at $Vd=0.0V$ for two fin heights. Device with higher fin height shows a greater buildup of carriers and therefore higher inversion capacitance. The lowering of threshold voltage with increased fin height is also observed from these figures. Fig. 11a shows the Id-Vg characteristics of the device at three different fin heights at $Vd=0.5V$ in both linear and log scale. Device with higher fin height results in higher carrier accumulation and therefore higher drain current. From the figure it is observed that the rate of drain current increment gradually drops as we move form *10nm* fin height to *14nm* fin height. This reveals the effect of the top gate becoming trivial in device operation as we increase fin height. Fig. 11b shows the effect of square channel cross-section on Id-Vg characteristics. Increased channel cross-section leads to higher drain current due to higher carrier accumulation. Subthreshold performance, threshold roll-off and short channel performance get degraded also with channel cross-section increment as seen from our simulation. Fig. 12a shows the effect of fin height on threshold voltage of the device. Device with higher fin height shows lower threshold voltage. This trend is consistent at all mole fractions of In used in this study.  Fig. 12b shows the effect of device dimension on subthreshold and short channel performance. Increasing fin height degrades short channel and subthreshold performance. These are evident from the increment in Subthreshold Swing (SS) and Drain Induced Barrier Lowering (DIBL) with increased fin height. Fig. 12c shows the effect of square fin cross-section on threshold voltage of the device. Lowering fin cross-section increases subband energies and therefore we see an increase in the threshold voltage of the device. Fig. 12d shows the effect of channel cross-section on SS and DIBL parameters. As seen from the figure, decreasing channel cross-section results in improved electrostatic control, better subthreshold and short channel performance. In these simulations, we have extracted SS and DIBL values for Id-Vg characteristics at $Vd=0.5V$. We have considered In mole fraction to be fixed at 0.7 for the study of SS and DIBL.

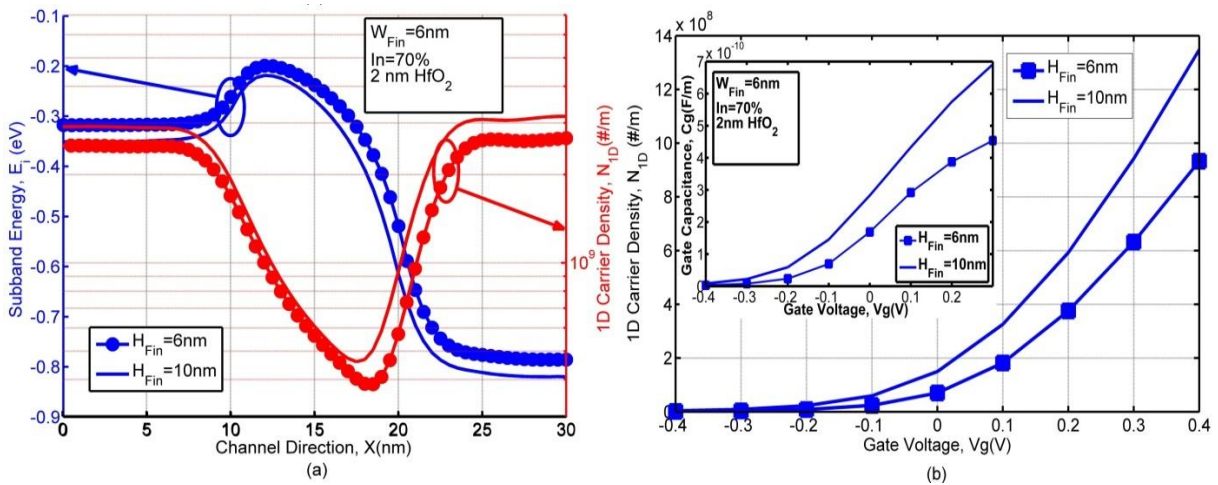

Fig. 10. (a) 1st subband energies and carrier density in the channel at two fin height conditions. The fin width is kept at *6nm*. Here, $Vg=0.4V$ and $Vd=0.5V$. Device with higher fin height results in higher carrier accumulation in the channel. (b) 1D carrier density at $Vd=0.0V$ at two different fin heights. The inset figure shows gate capacitance at different fin heights. Device with higher cross-sectional fin area shows higher carrier density, higher gate capacitance and lower threshold voltage.

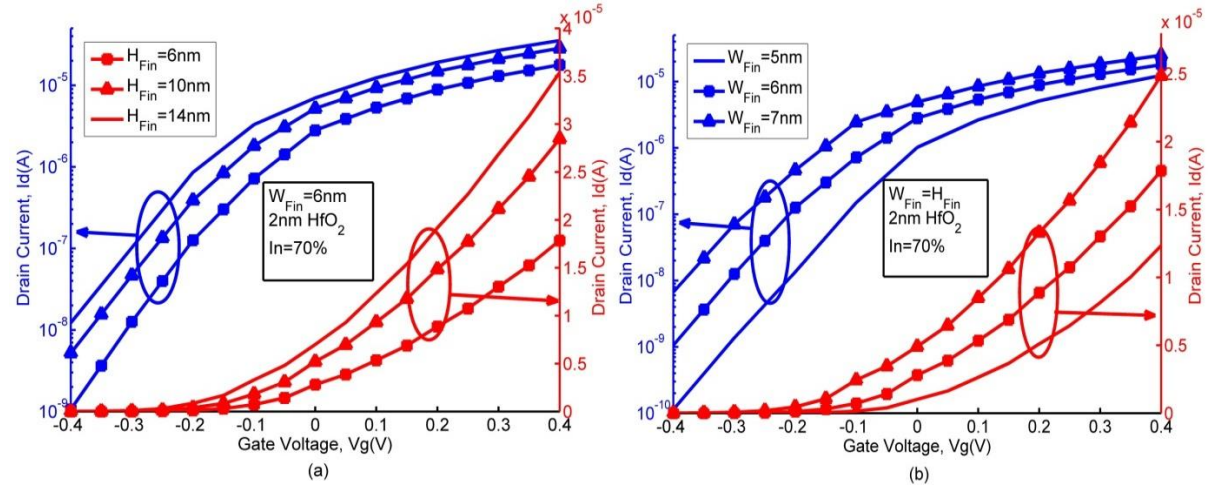

Fig. 11. (a) Id-Vg characteristics in both linear and log scale for various fin heights $V_d=0.5V$. Increment in fin height results in higher carrier accumulation and higher current drive. Device with higher fin height shows degraded short channel and electrostatic performance. (b) Id-Vg characteristics at different fin width conditions keeping $W_{Fin}=H_{Fin}$. Increased cross-sectional area gives higher drain current and degrades short channel performance. Threshold voltage gets lowered with increasing cross-sectional area. In these cases, In mole fraction is kept at 0.7.

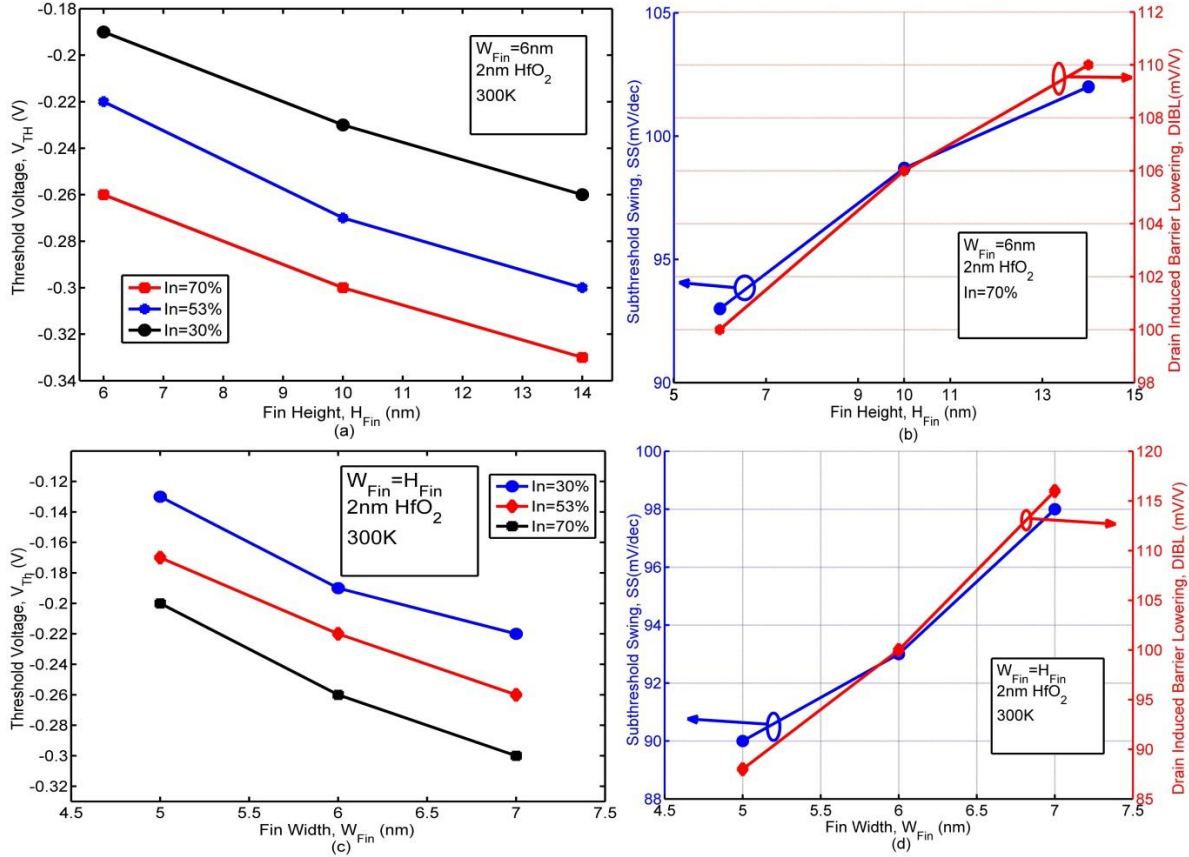

Fig. 12. (a) Variation of threshold voltage with fin heights at various In mole fractions in the channel. Increasing In mole fraction lowers energy bandgap and therefore lowers the threshold voltage of the device. (b) Variation of SS and DIBL with fin heights at In mole fraction of 0.7. (c) Threshold voltage variation at different channel cross-sections keeping $W_{Fin}=H_{Fin}$. (d) SS and DIBL values at different channel cross-sections keeping $W_{Fin}=H_{Fin}$. For, SS and DIBL studies, In composition in the channel is kept at 0.7.

DIBL refers to the modulation of energy barrier height with increased drain voltage. As devices are scaled, close proximity between source and drain makes device operation more prone to DIBL effects. From fig. 12b and fig. 12d, it can be seen that, deterioration of DIBL becomes more significant when we vary device cross-section by changing both fin width and height. This lowers the electrostatic effect of the gates on device operation. Which indicates contribution of 2D electrostatics effects in the deterioration of DIBL. On the other hand, the deterioration of SS can be attributed to mainly tunneling current which contributes significantly to drain current when device operation is in subthreshold regime.

**4.6 Effect of Channel Dimension, Gate Oxide Thickness on Carrier Mobility and Ballistic Injection Velocity:**

We have also explored the effect of device dimension on ballistic injection velocity of the carriers. The ballistic injection velocity can be extracted using eqn. 4.2.1 at device 'on' condition. Fig. 13a shows the effect of square channel cross-section on the ballistic injection velocity of the device. Fig. 13b shows the effect on fin height variation on the ballistic injection velocity while the fin width is kept fixed at 6nm. In both the cases we see an increment in injection velocity with device dimension. This can be explained by lowering subband energies and energy barrier height with increased device dimension. This phenomenon allows more source injected carriers in the channel and therefore we see an increase in the ballistic injection velocity with increased device dimension. Fig. 13c shows variation of ballistic injection velocity with top gate oxide thickness. As seen from the figure, Ballistic injection velocity gradually decreases with increased top gate oxide thickness and eventually shows a saturation tendency. This can be explained by movement of device operation to critical top gate oxide thickness limit that eventually leads to saturation in 'on' current profile as mentioned in sub-section 4.4. Fig. 13d shows variation of ballistic injection velocity with gate oxide thickness variation on all sides. With increased oxide thickness on all sides, the ballistic injection velocity appears to be decreasing which indicates decrease in device 'on' current with increased oxide thickness. In these figures, ballistic injection velocity at $Vg=0.4V$ and $Vd=0.5V$ has been shown. The variation of ballistic injection velocity with device channel dimension is consistent with the results observed in [9,21].

From the linear region of *Id-Vd* characteristics, we can extract ballistic carrier mobility in the channel. Carrier mobility for a non-planar device structure can be extracted from linear region of Id-Vd characteristics using the following relationship [17]:

$$I_d = \frac{\mu_B}{L_G} \cdot Q_i \cdot V_{DS} \tag{4.6.1}$$

In the study of mobility, $V_{DS}$ is considered to be *0.05V* where Id-Vd characteristics shows linear nature. Here, $Q_i$ refers to the carrier density at the top of the barrier.

Fig. 13e shows the variation of ballistic mobility in the channel as we vary 1D carrier density at the top of the barrier region. As the device moves into strong inversion region, we see a decrease in mobility as seen from the figure. Fig. 13f shows the variation of ballistic mobility with device dimension at strong inversion i.e. $Vg=0.4V$ and $Vd=0.05V$. Here, we have considered a square channel cross-section. With increased channel cross-section, we observe an increase in ballistic carrier mobility in the device. This increase can be attributed to increased drain current and higher carrier velocity at increased channel cross-section. The observed trend in ballistic mobility variation is consistent with the results shown in recent literature for InAs-OI FETs [9].

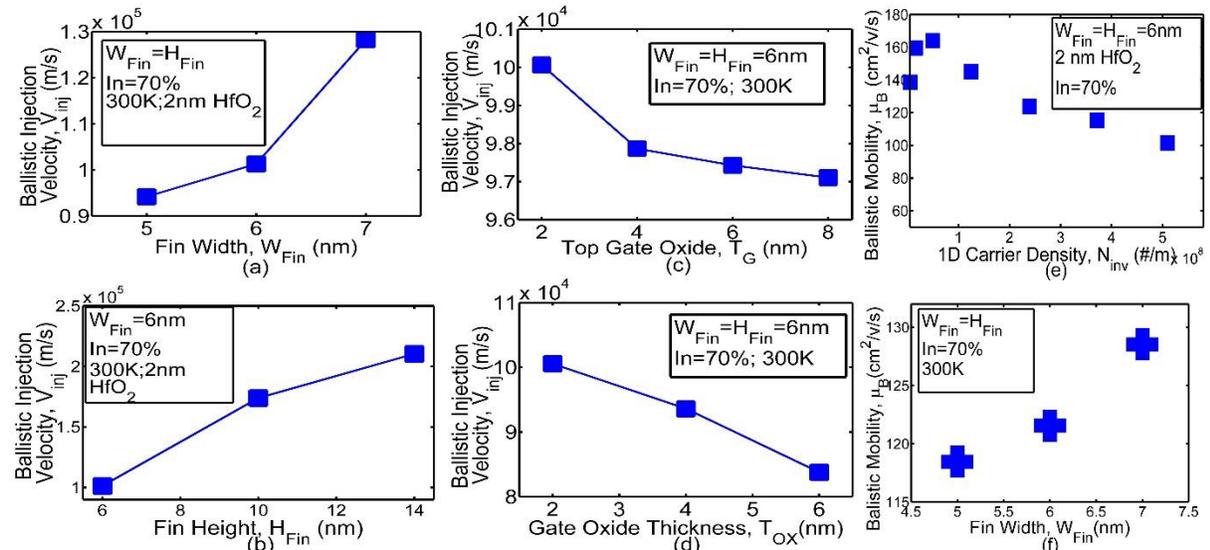

Fig. 13(a) Ballistic injection velocity at different fin dimensions keeping $W_{Fin}=H_{Fin}$. (b) Variation of ballistic injection velocity at different fin heights at $W_{Fin}$=6nm. (c) Variation of ballistic injection velocity at different top gate oxide thickness conditions. (d) Variation of ballistic injection velocity at different gate oxide thickness conditions while oxide thickness is varied in all directions. (e) 1D carrier density along with ballistic mobility for the device with square cross-sectional channel. (f) Variation of ballistic mobility with device dimension keeping $W_{Fin}=H_{Fin}$ at $Vg=0.4V$. Ballistic mobility appears to be increasing with increasing device cross-section.

## 5. Conclusion:

In this work we have presented a detailed quantum ballistic simulation study of a III-V Tri-gate MOSFET at *10nm* gate length. We have also observed and presented the effects of various physical device parameters like channel dimension, top gate oxide thickness, fin height, variable In composition in the channel on the transport characteristics of the device. Electrostatic performance of the device is also observed on the basis of quantum ballistic simulation results. Increasing In composition in the channel lowers threshold voltage of the device and improves ballistic current. Increasing fin height allows higher carrier accumulation in the channel and increases gate capacitance and device 'on' current as observed from the simulation. With increased fin height, the effect of the top gate becomes trivial gradually. However, the short channel and subthreshold performance of the device gets degraded with increased fin height. Increasing channel cross-section keeping $W_{Fin}=H_{Fin}$ lowers threshold voltage and increases device 'on' current. However, increased channel cross-section degrades short channel performance and subthreshold performance of the device. As for top gate oxide thickness, the ballistic simulation shows that, beyond *8nm* the increment of top gate oxide thickness does not change 'on' current significantly. Similar effect of top gate oxide thickness is also observed for carrier density in the device. Varying oxide thickness on all three gate directions, degrades electrostatics performance and degrades device 'on' current, carrier density and carrier velocity. Device dimension also plays a vital role in ballistic carrier velocity and effective carrier mobility in the channel. Increasing channel cross-section increases carrier injection velocity and ballistic mobility also increases. The results and observation of this study could help understand carrier transport in sub-*10nm* III-V multigate device structures.

**Acknowledgements:**

The authors are thankful to Md. Kawsar Alam for his valuable suggestions and fruitful discussions.

Kanak Datta                                                Department of Electrical Engineering

[20]   Caruso E, Lizzit D, Osgnach P, Esseni D, Palestri P, Selmi L. Simulation analysis of III-V n -MOSFETs : channel materials , Fermi level pinning and biaxial strain. Electron Devices Meet (IEDM), 2014 IEEE Int 2014:1–4. doi:10.1109/IEDM.2014.7047006.

[21]   Baek R, Kim D, Kim T, Shin CS, Park WK, Michalak T, et al. Electrostatics and Performance Benchmarking using all types of III-V Multi-gate FinFETs for sub 7nm Technology Node Logic Application. 2014 Symp VLSI Technol Dig Tech Pap 2014:168–9. doi:10.1109/VLSIT.2014.6894420.

[22]   Multiphysics Simulation Software - Platform for Physics-Based Modeling n.d. http://www.comsol.com/comsol-multiphysics.

[23]   Ren Z, Venugopal R, Goasguen S, Datta S, Lundstrom MS. nanoMOS 2.5: A two-dimensional simulator for quantum transport in double-gate MOSFETs. IEEE Trans Electron Devices 2003;50:1914–25. doi:10.1109/TED.2003.816524.

[24]   Kurniawan O, Bai P, Li E. Ballistic calculation of nonequilibrium Green's function in nanoscale devices using finite element method. J Phys D Appl Phys 2009;42:105109. doi:10.1088/0022-3727/42/10/105109.

[25]   Liu Y, Neophytou N, Low T, Klimeck G, Lundstrom MS. A tight-binding study of the ballistic injection velocity for ultrathin-body SOI MOSFETs. IEEE Trans Electron Devices 2008;55:866–71. doi:10.1109/TED.2007.915056.

[26]   Liu, Yang and Neophytou, Neophytos and Klimeck, Gerhard and Lundstrom MS. Band-Structure Effects on the Performance of III-V Ultrathin-Body SOI MOSFET s. Electron Devices, IEEE Trans 2008;55:1116–22.

[27]   Piprek J. Semiconductor optoelectronic devices: introduction to physics and simulation. Academic Press; 2003.

[28]   Van de Walle CG. Band lineups and deformation potentials in the model-solid theory. Phys Rev B 1989;39:1871.

[29]   Datta S. Nanoscale device modeling: the Green's function method. Superlattices Microstruct 2000;28:253–78. doi:10.1006/spmi.2000.0920.

[30]   Datta S. Quantum Transport : Atom to Transistor. Cambridge University Press; 2005. doi:10.1017/CBO9781139164313.

[31]   Rahman A. Exploring New Channel Materials for Nanoscale CMOS Devices: A Simulation Approach. PhD Thesis 2005.

[32]   Natori K. Ballistic metal-oxide-semiconductor field effect transistor. J Appl Phys 1994;76:4879–90. doi:10.1063/1.357263.

[33]   Lundstrom M, Ren Z. Essential physics of carrier transport in nanoscale MOSFETs. IEEE Trans Electron Devices 2002;49:133–41. doi:10.1109/16.974760.